\documentclass[aps,prb,twocolumn,showpacs]{revtex4-1}
\usepackage{graphicx}
\usepackage{amsmath,amsfonts}
\usepackage{color}

\newcommand{\ee}{{\rm e}}
\newcommand{\nR}{n_{\rm R}}
\newcommand{\nF}{n_{\rm F}}
\newcommand{\nsat}{n_{\rm sat}}
\newcommand{\gammac}{\gamma_{\rm C}}
\newcommand{\gammar}{\gamma_{\rm R}}
\newcommand{\gammaf}{\gamma_{\rm F}}
\newcommand{\gr}{g_{\rm R}}
\newcommand{\gf}{g_{\rm F}}
\newcommand{\gc}{g_{\rm C}}
\newcommand{\geff}{g_{\rm eff}}

\newcommand{\R}{R^{\rm 1D}}
\newcommand{\kap}{\kappa^{\rm 1D}}

\def\beq{\begin{equation}}
\def\eeq{\end{equation}}

\begin{document}

\title{Universality in nonequilibrium condensation of exciton-polaritons}

\author{Micha\l{} Matuszewski}
\affiliation{Instytut Fizyki Polskiej Akademii Nauk, Aleja Lotnik\'ow 32/46, 02-668 Warsaw, Poland}

\author{Emilia Witkowska}
\affiliation{Instytut Fizyki Polskiej Akademii Nauk, Aleja Lotnik\'ow 32/46, 02-668 Warsaw, Poland}

\begin{abstract}
We investigate the process of condensation of 
exciton-polaritons in a one-dimensional nanowire, predicting spontaneous formation of
domains of uncondensed excitons and condensed polaritons. We find that
the system does not follow  the standard Kibble-\.Zurek
scenario of defect formation. Nevertheless, the universal scaling laws
are still present, if the critical exponents are replaced by their ``nonequilibrium''
counterparts related to the imaginary part of the time-dependent Bogoliubov spectrum.
\end{abstract}

\maketitle

\section{Introduction}

Semiconductor microcavities, in which photons are confined and strongly coupled to excitons, allow for creation of 
exciton-polaritons. 
These bosonic quasi-particles can undergo Bose-Einstein condensation at standard cryogenic or even room 
temperature~\cite{Kasprzak_BEC,Nanowire,Polaritons,Wouters_Excitations,Deveaud_VortexDynamics}. 
The achievement of exciton-polariton condensate opened a path for new technological developments, 
eg.~in optoelectronics (ultra fast switches, quantum circuits)~\cite{Polariton_applications} or 
medicine (compact terahertz lasers)~\cite{terahertz lasers}.
It is known that condensation can have a character of 
a nonequilibrium phase transition when it takes place on a finite timescale~\cite{Damski_SolitonCreation}.
Such transitions have been attracting great attention in many fields of science~\cite{NonequilibriumBooks}. 
One of the early studies concerned topological defect formation 
in cosmological models~\cite{Kibble}. As the hot universe cooled down, the phase transitions
broke symmetries 
of the vacuum fields, which leaded to the creation of topological defects. This mechanism 
has been also considered in a wide variety of other physical models, including superfluid Helium~\cite{Helium_KZ}, superconductors, 
cold atomic gases, and other systems~\cite{other_KZ}. 
The outcome of the transition is the creation of defects such as
domain walls, monopoles, strings, vortices or solitons~\cite{Kibble,Helium_KZ,other_KZ,Damski_SolitonCreation}.
Bose-Einstein condensates of exciton polaritons, as highly controllable nonequilibrium 
quantum systems, are very suitable for studying these phenomena.

The Kibble-\.Zurek mechanism (KZM) theory is a powerful tool that allows to predict the density of defects 
without solving the full dynamical equations.
It is assumed that the system adiabatically follows the equilibrium state up to the ``freeze-out'' time $\hat{t}$ before
the critical point. At this moment the relaxation time becomes too long for the system to adjust to the change
of parameters. Consequently, the fluctuations approximately freeze, and the system enters the ``impulse'' phase.
After crossing the critical point, the adiabaticity is restored when the relaxation time becomes short enough again.
If the system in the final phase exhibits symmetry breaking, distant parts will choose to break the symmetry in
different ways. The number of resulting defects will be determined by 
the ``memorized'' healing length $\hat{\xi}$ at the time $\hat{t}$. 
Remarkably, the time $\hat{t}$ and the number of defects $N_{\rm d}$ can be determined solely from the 
critical exponents of the phase transition,
according to the scaling laws
\begin{equation} \label{scaling}
|\hat{t}| \sim v^{-z\nu/(1+z\nu)}\,, \quad N_{\rm d} \sim \hat{\xi}^{-d} \sim v^{d\nu/(1+z\nu)}\,,
\end{equation}
where $v$ is the rate at which the critical point is crossed, $d$ is the dimensionality, $z$ and $\nu$ are the critical exponents.
All the details of complicated dynamics are hidden, but we are still able to derive Eqs.~(\ref{scaling})
thanks to the underlying universality.
The Kibble-\.Zurek scaling laws have been confirmed numerically in many cases, including 
the condensation of ultracold atoms~\cite{Damski_SolitonCreation}.

The condensation of exciton-polaritons, 
on the other hand, is an example of 
a qualitatively different case. This
phase transition may be named ``nonequilibrium'' in yet another meaning of this word~\cite{Wouters_Excitations,NonequilibriumPolaritons}. 
In fact, there may not be a well defined equilibrium state that the system could follow, due to the lossy nature
of the system itself~\cite{NonequilibriumBooks}.

A natural question arises, whether the Kibble-\.Zurek scenario does apply to the case of an intrinsically nonequilibrium 
system such as the polariton condensate? And do the critical scaling laws such as the one given 
by Eq.~(\ref{scaling}) hold? 
In this paper, we aim to answer these questions by investigating exciton-polariton
condensation in a one-dimensional GaAs nanowire~\cite{Nanowire}. We note that recently it was suggested that the KZM  
describes the formation of vortices in a related two-dimensional system~\cite{Deveaud_VortexDynamics}. However,
no evidence of the above adiabatic-impulse-adiabatic scenario or the critical scalings was given.
We show that the dynamics cannot be adequately described by the standard KZM, but 
scalings similar to Eq.~(\ref{scaling}) are still present.

\section{The model}

We model polariton condensation in a nonresonantly pumped nanowire within the truncated Wigner approximation adapted 
to the polariton case~\cite{ClassicalFields,Deveaud_VortexDynamics}. The stochastic equations for
the polariton order parameter $\psi(x,t)$ and the densities of the exciton reservoir $\nR(x,t)$ and free carriers 
$\nF(x,t)$ are 
(in the one-dimensional version)
\begin{align} \label{GP_psi}
d \psi &= -\frac{i}{\hbar} dt \bigg[-\frac{\hbar^2}{2 m^*} \frac{d^2}{dx^2}
+ U(x)  + i\frac{\hbar}{2}\left(\R \nR  - \gammac \right) +\nonumber \\
&+\left( \gc^{\rm 1D} |\psi|^2 + \gr^{\rm 1D} \nR + \gf^{\rm 1D} \nF \right)  \bigg] \psi + dW\,,\nonumber\\
\frac{\partial \nR}{\partial t} &= \kap \nF^2 - \gammar \nR - \R \nR |\psi|^2\,,\\
\frac{\partial \nF}{\partial t} &= P(x,t) -  \kap \nF^2 - \gammaf \nF\,\nonumber,
\end{align}
where $m^*$ is the effective mass of lower polaritons, $U(x)$ is the disorder potential due to 
imperfections of the sample,
$R$ is the rate of stimulated scattering, $\gammac$, $\gammar$, and $\gammaf$ are the polariton, 
exciton, and free carrier loss rates,
$\gc$,  $\gr$, and $\gf$ are the respective interaction coefficients, $\kappa$ is the exciton formation 
rate~\cite{Szczytko_ExcitonFormation},
$P(x,t)$ is the free carrier creation rate,
and $dW$ is a complex stochastic variable with
\begin{align} \label{dW}
\langle dW(x) dW(x')\rangle &= 0\,, \\ \nonumber
\langle dW(x) dW^*(x')\rangle &= \frac{dt}{2 \Delta x} (\R \nR + \gamma_C) \delta_{x,x'}\,,
\end{align}
reflecting the quantum noise due to the particles entering and leaving the condensate.
The coefficients $R$, $\kappa$, $\gc$, $\gr$, and $\gf$ are rescaled in the 
one-dimensional case by assuming a Gaussian
perpendicular profile of $|\psi|^2$, $\nR$ and $\nF$ of width $d$ determined by the nanowire thickness, 
$(\R,\kap,g_i^{\rm 1D})=(R,\kappa,g_i)/\sqrt{2\pi d^2}$.

An ensemble of solutions of these stochastic equations follows the evolution of the Wigner function
of the system in phase space. They allow to compute various correlation functions when averaging over the ensemble, 
and a single solution can be interpreted as a possible result of a single experiment (single pulse). 
Below, we take advantage of both of these interpretations.
The above model describes the dynamics of the system at zero temperature. In reality, the distribution
of excited modes may resemble a thermal state~\cite{Kavokin_Microcavities,Kasprzak_BEC,ClassicalFields}. 
However, such distribution should not influence the 
phenomena presented here. We will see that the size of domains
is equal to several tens of micrometers, 
one order of magnitude larger than the typical correlation (de Broglie) length in a thermal 
state at 10 K. Consequently, such state can be
approximated by a flat distribution in momentum space, in accordance with Eqs.~(\ref{dW}).

\section{Creation of exciton and polariton domains}

We consider an excitation pulse with a Gaussian temporal profile and a flat spatial profile of extent $L$
\beq
P(x,t)= P_0 \exp(-(t/t_0)^2) \theta(L/2-|x|), 
\eeq
with $\theta(x)$ being the Heaviside step function and $t_0=200\,$ps.
The general conclusions should hold for other spatial and temporal 
pulse shapes, as well as electrical pumping, provided that the pump is switched on rapidly enough.
We checked that the same phenomena occur even in the case of an infinitely short $\delta(t)$-pulse.

\begin{figure}
\includegraphics[width=8.5cm]{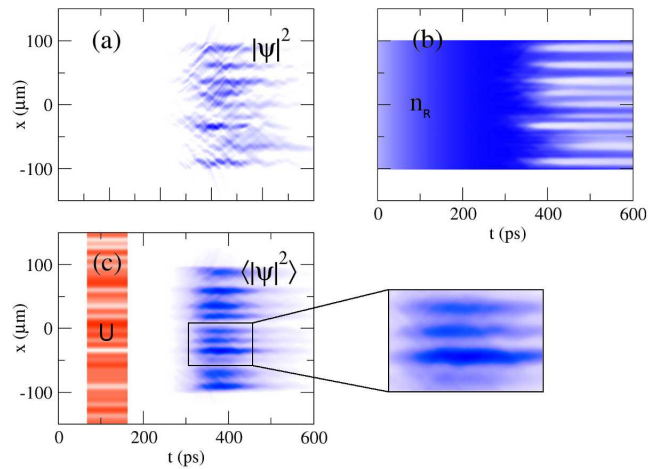}
\caption{Evolution of (a) polariton density $|\psi|^2$ and (b) reservoir exciton density $\nR$ during condensation
in a 1D wire excited by a pulse with peak intensity $P_0=7.7 P_{\rm th}$
and duration $t_0=200\,$ps. 
(c) Polariton density averaged over 100 pulses. The disorder potential $U(x)$ of average amplitude $U_0=0.05\,$meV used in (a)-(c) 
is depicted in (c) on the left.
Other parameters are $\gammac=1/30\,$ps, $\gammar=1/400\,$ps, 
$\gammaf=1/3\,$ns, $\kappa=10^{-4} \mu$m$^2\,$ps$^{-1}$,
 $d=5 \mu$m, $R=2.8\times 10^{-3} \mu$m$^2$ ps$^{-1}$, $\gc=3 \mu$eV $\mu$m$^2$, $\gr = \gf= 2\gc$.}
\label{fig:frames}
\end{figure}

In Fig.~\ref{fig:frames} we present results of a typical simulation of the condensation process. 
The Figures~\ref{fig:frames}(a) and~(b) show the polariton and reservoir exciton density in a single simulation,
which can be interpreted as a single experimental realization (pulse)~\cite{TruncatedWigner}. 
The polariton condensate appears 
about $300\,$ps after arrival of the pulse,
in the form of areas of high density, complemented by areas of high reservoir exciton density. 
While the polariton domains appear to be vulnerable and erratic structures, 
exciton domains are much clearer due to the large exciton diffusion time.
Here we included a realistic disorder potential $U(x)$ with amplitude of a fraction of meV and correlation length of a few micrometers. 
We checked that the only visible effect of such disorder is to introduce preferential positions for the domains in the potential minima.
The domain widths or other characteristics are not changed substantially if the disorder
is removed from the model. Thus, the domain creation phenomenon is not directly caused by the potential, 
but related to symmetry breaking as in the Kibble-\.Zurek mechanism.

The polariton structures become much better defined when
the density is averaged over many pulses, see Fig.~\ref{fig:frames}(c). 
The disorder allows to pin the domains in certain regions of the sample, that
would otherwise become smeared out over many pulses (see Appendix~\ref{sec:realizations}). 
The apparent phase separation originates 
from the repulsion between polaritons and excitons. It is related to the existence
of two stationary solutions to the homogeneous version of Eqs.~(\ref{GP_psi}).
Substituting  $\psi(x,t) = \psi_0 \ee^{-i \mu t}$, $\nR(x,t)=n_{\rm R 0}$ and $\nF(x,t)=n_{\rm F 0}$ we get
\begin{align}
(a)\quad &|\psi_0|^2 = 0, &n_{\rm R 0}&=\frac{P_0}{\gammar}, \\
(b)\quad &|\psi_0|^2 = \left(\frac{P_0}{P_{\rm th}} - 1\right)\frac{\gammar}{\R}, &n_{\rm R 0}&=n_{\rm th} \equiv\frac{\gammac}{\R}, \nonumber
\end{align}
where $P_{\rm th}=\gammac\gammar/\R$ is the threshold value of the pumping intensity $P_0$.
Here we assumed that $\gammaf^2 / \kappa \ll \gammac \gammar  / \R$, which is a good approximation for the parameters
used.
The above (a) and (b) are the exciton and polariton phases, the latter existing only for $P_0>P_{\rm th}$.

The presence of the disorder adds another complication, as the polariton condensate has the tendency 
to be localized in its minima. We note that creation of domains through symmetry breaking 
and simple localization can, in principle, 
coexist in the same experiment -- defect formation during the phase transition could be followed 
by localization in the potential minima at late stages of evolution. The question is whether the 
simple localization could mimic the defect formation, and how to distinguish between the two. 
We show that this can be easily achieved by investigating the dependence of the number of 
domains in function of the pumping intensity. 

\begin{figure}
\includegraphics[width=8.5cm]{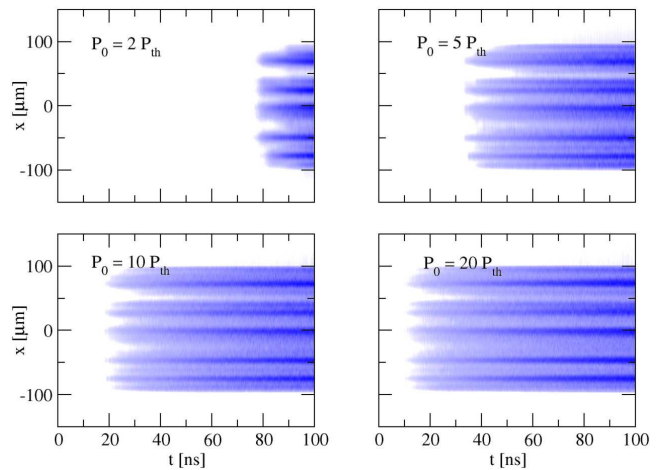}
\caption{Localization of polariton density in the minima of a disorder potential. 
Here, the pumping intensity is turned on slowly, reaching the value $P_0$ at $t = 100\,$ ns,
and other parameters are as in Fig.~\ref{fig:frames}. 
The positions of domains
correspond to the potential minima, and their number does not depend on the pumping intensity.}
\label{fig:localization}
\end{figure}

\begin{figure}
\includegraphics[width=8.5cm]{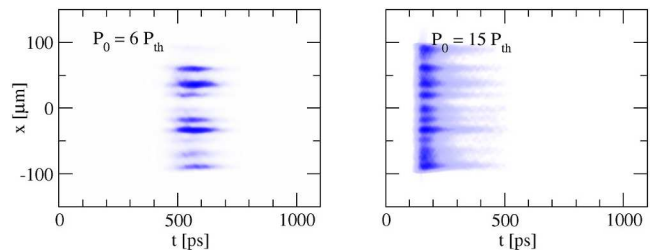}
\caption{Creation of domains by a short Gaussian pulse of duration $t_0=200\,$ ps.
The number of domains depends on the peak pumping intensity $P_0$, as shown in Fig.~4 in the main text.
The number of counted domains~\cite{counting_fluctuations} is 11 (5 polariton + 6 exciton) 
for $P_0=6P_{th}$ and 21 (10 polariton + 11 exciton) for $P_0=15P_{th}$.
The density is averaged over 100 pulses.
Other parameters as in Fig.~\ref{fig:frames}.}
\label{fig:f2}
\end{figure}

We consider a scenario (or a ``control experiment''), where the pumping intensity is increased slowly.
Such transition can be achieved eg.~by using a continuous wave laser beam for pumping.
This corresponds to a very slow phase transition, and a very large domain length scale $\xi^*$ in our model,
larger than the system size L (we use switching time equal to 100ns which is long concerning 
the time scales of the system). In this case, polariton domains, or ``condensates localized in 
the potential minima'', are still created, but their number is practically independent on the 
pumping intensity, see Fig.~\ref{fig:localization}.
The domains always appear in the potential minima,
and there is no scaling law for the number of domains. This is in contrast with the case of excitation by a picosecond pulse, where 
the number of domains strongly depends on the peak intensity, see Fig.~\ref{fig:f2}. 
We note that in this case the number of domains is never lower than the number 
of domains resulting from the localization in the potential minima.

\section{Analysis of the domain formation process}

\begin{figure}
\centerline{\includegraphics[width=8.5cm]{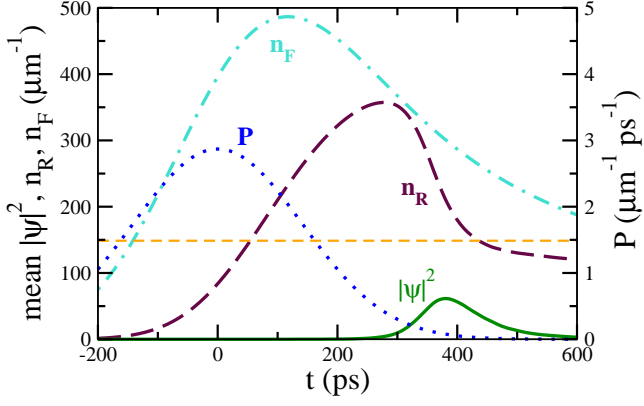}}
\caption{Evolution of the mean polariton density $|\psi|^2$, reservoir exciton density $\nR$, and
free carrier density $\nF$, in Fig.~\ref{fig:frames}
averaged over the illuminated area. The dotted line shows the pump intensity $P(x=0,t)$. 
The horizontal line corresponds to the threshold value
of $\nR=n_{\rm th}$ when the polariton field becomes unstable, $\epsilon=0$. 
}
\label{fig:dynamics}
\end{figure}

Let us turn our attention to the early dynamics of the system. 
As shown on Fig.~\ref{fig:dynamics}, after arrival of the pulse,
first the free carrier density $\nF$, and subsequently
the reservoir exciton density $\nR$ increase gradually. 
The polariton field, on the other hand,
is damped or amplified depending on the value of 
\beq
\epsilon(x,t)=\R \nR - \gamma_C=\R (\nR - n_{\rm th}).
\eeq
The condition $\epsilon=0$ determines the critical point, and $\epsilon$ plays the role
of the relative temperature~\cite{Helium_KZ}. 
We are not able to control $\epsilon$ directly, but through $P_0$ 
we can change the slope of $\epsilon(t)$ at the critical point.

The crucial observation is that while $|\psi|^2$ remains small, the 
system is in a strongly nonequilibrium state, with the related timescale 
given by $t_{\rm rel}=1/|\epsilon|$. There is practically no mixing of modes with different momenta, 
and the relaxation does not establish any 
relevant length scale. In other words, the weakness of interactions doesn't allow for the
effective thermalization that could lead to a quasi-equilibrium and the appearance of  a condensed state.
Contrary to the KZM picture, there is no well defined equilibrium before the phase transition, 
and the polariton mode occupation is shaped by the noise correlations~(\ref{dW}). Moreover, as we show
below, it is not these initial correlations that determine the final number of created defects.

The situation changes when the polariton-polariton interactions become sufficiently strong
to compete with nonequilibrium processes, $\epsilon(\tilde{t}) = \chi \gc^{\rm 1D} \overline{n}(\tilde{t})$,
with $\chi$ being a constant of the order of unity, and $\overline{n}(t)=N(t)/L$ the average polariton 
density in the illuminated area. 
At this point  $\tilde{t}$, which can be identified as the condensation time~\cite{Kavokin_Microcavities},
mode mixing starts to take place, which results in the appearance
of an instantaneous quasi-equilibrium.
The length scale $\xi^*$ associated with the emerging domains is established.
The number of domains is determined by  
\begin{equation} \label{eq:Nd}
N_d\sim L/\xi^*\,.
\end{equation}

In the following, we estimate the emerging length scale $\xi^*$.
We consider the dynamics of the system from the critical point $\epsilon=0$ crossed at $t=t_{\rm th}$
up to the time when the condensate density $|\psi|^2$ reaches its maximum, $t = t_{\rm max}$, see Fig.~\ref{fig:dynamics}.
We are interested in the appearance of a length scale during the growth of the polariton field
$\psi$. At a very short time scale, much smaller than $\gammac^{-1}$, $\gammar^{-1}$, 
and $(\kap \nF^2)^{-1}$, the total exciton density 
$n_{\rm ex} \equiv \nR + |\psi|^2$ is a constant of motion of Eqs.~(\ref{GP_psi}). 
This allows to derive an effective equation, which in the case $U=0$ reads
\beq \label{modelA}
i \hbar \dot {\psi} = \bigg[-\frac{\hbar^2 \nabla^2}{2 m^*}
+ i\frac{\hbar\gamma_0}{2} \left(1-\frac{|\psi|^2}{\nsat}\right) 
+\geff |\psi|^2 + \mu_{\rm a}  \bigg] \psi\,, 
\eeq
where $\geff=\gc^{\rm 1D} - \gr^{\rm 1D}$, $\mu_{\rm a}=\gr^{\rm 1D} n_{\rm ex}$, 
$\gamma_0 = \R (n_{\rm ex} - n_{\rm th})$, and 
$\nsat=n_{\rm ex} - n_{\rm th}$. We note that equation of the same form has been used by other authors 
to describe the polariton dynamics~\cite{SingleEquation}, with an important difference that the sign of 
the effective interaction $\geff$ is {\it negative} in the physically relevant case $\gc<\gr$. Obviously,
the above equation cannot model the long-time dynamics of repulsive particles, but is useful to describe dynamical 
instability on a {\it short} time scale. 

\begin{figure}
\centerline{\includegraphics[width=8.5cm]{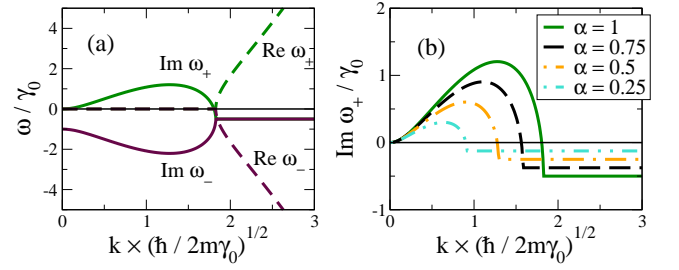}}
\caption{(a) Real and imaginary parts of Bogoliubov frequencies $\omega_\pm$ for a solution 
at the point of saturation, $\alpha=|\psi_0|^2/\nsat=1$. Note that the vertical and horizontal axes are rescaled by 
$\gamma_0^{-1}$ and $\gamma_0^{1/2}$. (b) Imaginary part of the unstable branch $\omega_+$ for several 
values of the saturation parameter $\alpha$.}
\label{fig:lambda}
\end{figure}

Let us now assume that there exists a patch of approximately constant polariton field $\psi_0$ (a condensate seed),
at $t=t_1$. We can describe small fluctuations around $\psi_0$ by 
\beq \label{Bogoliubov}
\psi = \left(\psi_0 + u(t) \ee^{ikx} + v^*(t) \ee^{-ikx} \right) \ee^{-i\mu t + \lambda (t-t_1)}\,,
\eeq
where $\mu$ and $\lambda=(\gamma_0/2)(1-|\psi_0|^2/\nsat)$ are the chemical potential and 
the growth rate of the patch, and $u$, $v$ represent the fluctuations.
Substituting the above to (\ref{modelA}) gives the Bogoliubov-de Gennes modes $u,v^* \sim \ee^{-i\omega_\pm t}$, 
and the mode frequencies
\beq
\frac{\omega_\pm}{\gamma_0} = -i\frac{\alpha}{2} \pm i\sqrt{\left(\frac{\alpha}{2}\right)^2 +(\alpha \beta)^2 - \left(
\varepsilon_k+\alpha \beta \right)^2}
\eeq
where $\varepsilon_k=\hbar k^2 / 2 m^* \gamma_0$, the saturation parameter $\alpha=|\psi_0|^2 / \nsat$, 
and $\beta=\geff/\hbar \R \approx -1.63$ for our parameters. 
As shown in Fig.~\ref{fig:lambda}, mode frequencies are complex in general,
with unstable long-wave modes (Im $\omega_+>0$)  below a certain value of $k_{\rm cutoff}\sim\gamma_0^{-1}$. 
If the patch is larger than the length scale set by $k_{\rm cutoff}^{-1}$, dynamical instability will 
lead to the breakup or shrinkage of the patch. In the opposite case, the patch will expand due to 
damping of short-wavelength perturbations. This ''spectral filtering'' is expected to produce structures
with a well defined length scale $k_{\rm cutoff}^{-1}$.
As shown in Fig.~\ref{fig:lambda}(b), this process is efficient for large values of the saturation parameter $\alpha$, 
when $|\psi_0|^2$ is
comparable to $\nR$. 
Such high polariton density is achieved when $\nR(t)$ reaches its peak value, i.e.~when
$\R n_{\rm ex} - \gammac \approx \epsilon_{\rm max}$, where $\epsilon_{\rm max}$ is the maximum value of 
$\epsilon(x,t)$ (see Fig.~\ref{fig:dynamics}). This allows to estimate the length scale
as $\xi^*= (\hbar / 2 m^* \epsilon_{\rm max})^{1/2}$.

\begin{figure}
\centerline{\includegraphics[width=8.5cm]{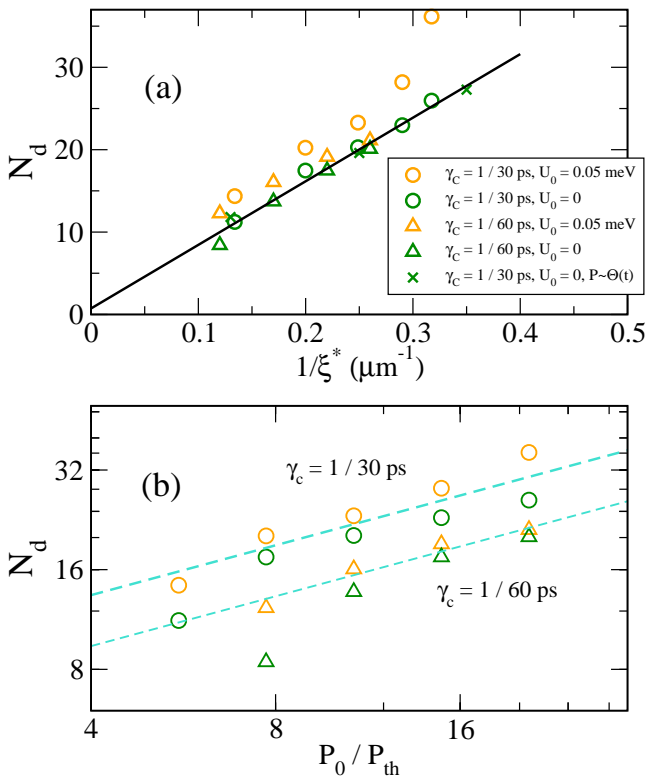}}
\caption{
Average number of  domains as a function of (a) $\xi^* = (\hbar / 2 m^* \epsilon_{\rm max})^{1/2}$
and (b) the normalized pulse intensity.
Note that the disorder-free $U_0=0$ averages can be only observed in single-pulse measurements.
Crosses in (a) correspond to simulations where the system was excited by a rapidly 
turned on laser beam $P(x,t)= P_0 \theta(t) \theta(L/2-|x|)$ rather than a Gaussian pulse.
The solid line is a linear fit to the data without disorder.
The dashed lines in (b) are $P_0^{1/2}$ guides to the eye, see~(\ref{scaling_law}).
}\label{fig:scaling-num}
\end{figure}

In Figure~\ref{fig:scaling-num} we plot the average number of created domains~\cite{counting_fluctuations}, 
according to numerical solutions of Eqs.~(\ref{GP_psi}) and~(\ref{dW}). 
In the cases with the disorder $U_0>0$, we average the result over 30 random disorder potentials of the same amplitude $U_0$.
For each of the disorder potentials, we run 50 simulations of the experiment with random Wigner noise and count the number of
domains present on the average density profiles similar to Fig.~\ref{fig:frames}(c).
We also show the results of disorder-free cases, where the averaging is performed only over random Wigner noise realizations,
and the number of domains is calculated using single-pulse pictures like Fig.~\ref{fig:frames}(a),(b).
We stress that in the latter case, domain counting could be preformed only in single-pulse measurements
or through $g^{(2)}$ correlation measurements, since the average density profile contains no domains due to lack of pinning.

In the absence of disorder, the number of domains is clearly inversely proportional
to $\xi^*=(\hbar / 2 m^* \epsilon_{\rm max})^{1/2}$ over the pumping range of $P_0=6 P_{\rm th} \dots 20 P_{\rm th}$.
The scaling prefactor appears to be exactly the same for two different values of $\gammac$
and even for a qualitatively different temporal $\theta(t)$ excitation.
This provides a robust confirmation of our model. Figure ~\ref{fig:scaling-num}(b)
shows the predicted $N_d\sim P_0^{1/2}$ scaling of the approximate analytical model, see Appendix~\ref{sec:analytical}.

\begin{figure}
\centerline{\includegraphics[width=8.5cm]{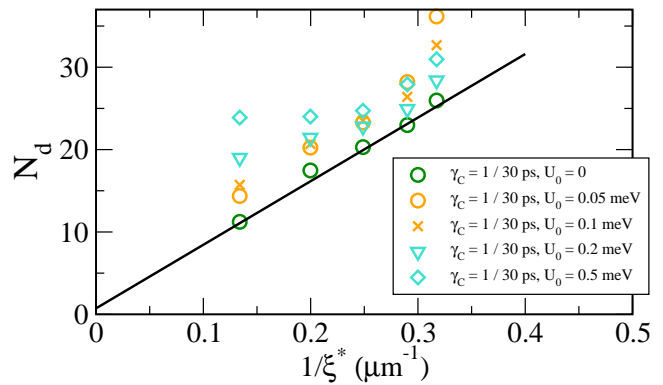}}
\caption{Same as in Fig.~\ref{fig:scaling-num}(a), for five different values of the disorder amplitude.
}\label{fig:scaling-VA}
\end{figure}

The presence of disorder has an important effect on the dynamics, as it is able ``pin'' the
domains to certain positions in the sample, see Appendix~\ref{sec:realizations}. This can be understood from the fact
that the initial density fluctuations are suppressed in the areas of higher potential $V(x)$.
Hence, it is more likely that condensate seeds will form in potential wells rather than hills.
In the momentum space, this effect results in redistribution of $\psi(k)$ mode occupation.
In the lowest order approximation, this effect will be washed out in $k$ space by the consequent spectral filtering,
leaving no fingerprint on the density of created domains. The fast exponential growth of selected modes
would make any initial spectral mode occupation irrelevant. At the same time, domain positions in real space would 
essentially remain unaffected by the growth.
Such approximation, however, doesn't take into account the influence of the potential on the filtering process.
In fact, in simulations with realistic disorder amplitude, we can see an increase of the number of domains with respect to
the disorder-free case. 
This is especially visible for larger values of the disorder amplitude, 
as shown in Fig.~\ref{fig:scaling-VA}. For the largest amplitude,
$V_0=0.5\,$meV, there is clearly a lower limit for the number of domains, which we identify with the
the number of domains in the stationary state of the disorder potential, or
the number of stationary condensates.
This natural lower limit for $N_d$ may preclude the observation of scaling in 
samples with large amount of disorder.
For the largest pumping intensity, the number of domains (light circles) may also deviate from the scaling law, which is associated
with the instability of domains for strongest pumping and the associated difficulty in domain counting. 
This is also visible in large fluctuations of the domain number.

\section{Discussion}

It is clear that some of the key elements of the KZM, such as the 
adiabatic-impulse-adiabatic scenario, or the ``freezing'' of correlation length before the phase transition, are 
not present in the condensation dynamics. 
Nevertheless, the scaling laws also result from the competition
of timescales existing in the system. 
To obtain them, let us define ``nonequilibrium'' counterparts of critical exponents $z$ and $\nu$ with
an analogue to the inverse correlation length, $k_{\rm cutoff} \sim \epsilon^{\tilde{\nu}}$, and an analogue to the inverse 
relaxation time, $\lambda \sim \epsilon^{\tilde{z}\tilde{\nu}}$.
If we treat $\epsilon$ as an external control parameter that
changes like $\epsilon \sim v(t-t_{\rm th})$, the local departure 
of the order parameter from the initial ``vacuum'' state is proportional 
to $|\psi|^2\sim \exp(v^{\tilde{z}\tilde{\nu}} (t-t_{\rm th})^{1+\tilde{z}\tilde{\nu}})$, 
and the quasi-equilibrium is achieved
at $\tilde{t} - t_{\rm th} \sim v^{-\tilde{z}\tilde{\nu}/(1+\tilde{z}\tilde{\nu})}$, up to a logarithmic factor. 
The spectral cutoff at this point is $\tilde{k}_{\rm cutoff} \sim v^{-\tilde{\nu}/(1+\tilde{z}\tilde{\nu})}$, which gives the 
defect number $N_d\sim v^{d\tilde{\nu}/(1+\tilde{z}\tilde{\nu})}$.
Remarkably, these formulas have identical forms as the standard KZM scalings~(\ref{scaling}). For example, in the present system
we have $\tilde{z}\tilde{\nu}=1$ and $\tilde{\nu}=1/2$, which gives the correct 
prediction $N_d\sim v^{1/4} \sim P_0^{1/2}$, see~(\ref{scaling_law}). Note that while the mechanism of spectral filtering is somewhat 
similar to phase ordering kinetics~\cite{Bray}, contrary to the latter, it describes the stage of defect formation, 
and may be used to provide a starting point for the subsequent phase ordering.

\section{Conclusions}

In conclusion, we investigated the formation of coherence of 
exciton-polariton condensate in a one-dimensional nanowire.
We demonstrated the spontaneous formation of exciton and polariton domains and found 
that the standard adiabatic-impulse-adiabatic scenario of the Kibble-\.Zurek theory cannot describe the system dynamics.
Nevertheless, appropriately modified scaling laws are still present, which results from the competition 
between characteristic timescales in the system.
Our study can have practical implications for applications proposed for polariton 
systems, as we point out the possible difficulty in creating a coherent condensate 
on a short timescale. 

\acknowledgments

We thank Jacek Dziarmaga for reading the manuscript and continuous stimulating discussions.
This work was supported by the National Science Center grants DEC-2011/01/D/ST3/00482, DEC-2011/03/D/ST2/01938,
and by the Foundation for Polish Science through the \textquotedblleft Homing Plus\textquotedblright\ program. 

\appendix 

\section{Averaging over multiple pulses}
\label{sec:realizations}

\begin{figure}[h]
\includegraphics[width=8.5cm]{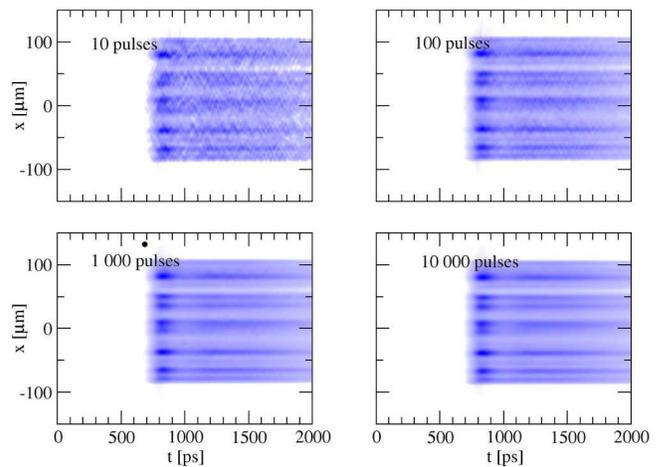}
\caption{Example of polariton density patterns averaged over many pulses. The weak disorder potential ``pins''
the domains to certain preferred positions, which results in clear patterns even after averaging over thousand of pulses.
Parameters are the same as in Fig.~1 in the main text, 
except $P_0=5 P_{\rm th}$ and pulse length $t_0=1000\,$ps.}
\label{fig:frame_realizations}
\end{figure}

While the disorder potential introduces some deviations from the scaling laws,
its presence is useful for the observation of domains in a streak-camera experiment.
In the absence of the disorder, the domains are created at random positions in the sample, and consequently the averaged 
luminescence intensity pattern becomes smeared out for a large number of pulses. Remarkably, the disorder potential
with appropriate amplitude has the ability to ``pin'' the domains to certain preferred positions, which allows
to obtain  clear patterns even when averaging over very many pulses. The Figure~\ref{fig:frame_realizations}
shows typical results of averaging. With the increase of the number of pulses the picture becomes clearer, while the intensity
contrast is not diminished. The picture suggest 
that the picture will not change even after averaging over millions of pulses as in real experiments.

\section{Effect of the polariton energy relaxation}

\begin{figure}
\includegraphics[width=8.5cm]{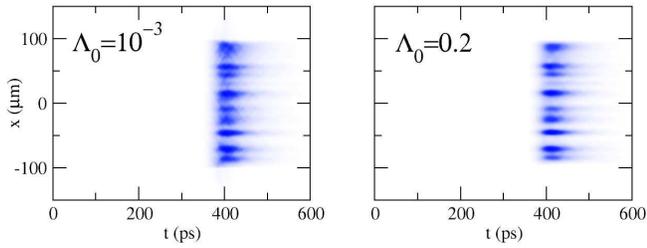}
\caption{An example showing the effect of the polariton energy relaxation for two values 
of $\Lambda_0$. Here $P_0=10 P_{\rm th}$.
The density is averaged over 100 pulses.
Other parameters as in Fig.~\ref{fig:frames}.}
\label{fig:relaxation}
\end{figure}

Several experiments~\cite{Nanowire,1DAmplification,GapStates} have shown that the polariton 
energy relaxation (or thermalization) 
may play an important role in their dynamics. To estimate the effect of relaxation due to the interactions
with the exciton reservoir, which seems to be the dominant mechanism when the spatial overlap between the reservoir
and the condensate is significant, we follow the phenomenological model of~\cite{Nanowire,1DAmplification,GapStates}
by adding the additional term of the form 
$+\Lambda_0 \frac{\nR(x,t)}{\nR^{max}} \frac{\hbar}{2m^*}\nabla^2\psi dt$ to the
right hand side of the polariton field equation. Here $\nR^{max}$ is the maximum density of exciton reservoir over 
the whole evolution, and the magnitude of the relaxation coefficient~\cite{1DAmplification,GapStates} $\Lambda_0$ ranges 
from $10^{-3}$ to $0.2$. We find that for $\Lambda_0=10^{-3}$ the effect
of relaxation on the domain formation is negligible, and even for $\Lambda_0=0.2$ it does not affect the 
domain pattern structure, see Fig.~\ref{fig:relaxation}. In the latter case, the domain pattern is smoothed out,
which is due to the additional damping of short-wave fluctuations.

\section{Defect formation vs. the hole-burning effect}

\begin{figure}[h]
\includegraphics[width=6.5cm]{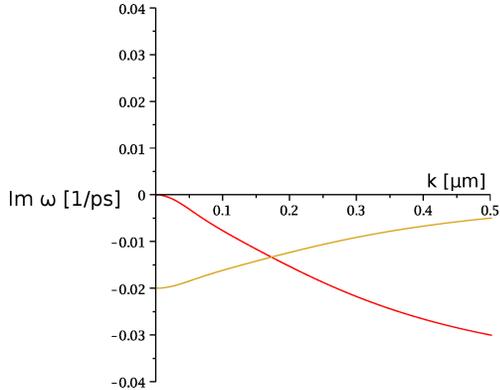}
\caption{Imaginary parts of the Bogoliubov eigenfrequencies obtained using the full model (2) instead of (7),
corresponding to the steady-state polariton condensate solution,
Eq.~(4b) in the main text. The brown line
corresponds to two modes with the same imaginary parts. Parameters are the same as in Fig.~1 in the main text, 
except $\gammac=1 / 60\,$ps and $P_0=15 P_{\rm th}$.}
\label{fig:static}
\end{figure}

Our model predicts the creation of defects due to the special form of the Bogoliubov excitation spectrum
during the buildup of the polariton phase. It was demonstrated~\cite{Wouters_Excitations} that
the excitation spectrum of a steady-state polariton condensate can also develop 
a similar band of unstable modes at at small $k$, which may give rise to a 'hole-burning' effect,
and formation of polariton and reservoir exciton domains. However, we found that this effect alone
cannot explain our results. In particular, domains are formed also when the steady-state 
excitation spectrum does not exhibit any instability, as shown in Fig.~\ref{fig:static}.
Consequently, the phenomenon we describe is a dynamical effect, and cannot be explained
by the analysis of a steady-state solution.

\label{sec:analytical}
\section{Analytical model of system dynamics}

When introducing our model we suggest that the number of domains is controlled by the rate $v$ at which the critical
point is crossed, in analogy to the theory of the Kibble-\.Zurek mechanism. As shown in Fig.~2 in the main text,
the reservoir exciton density $\nR$ grows approximately linearly at the critical point $\epsilon=0$, which allows
to make analytical predictions about the resulting length scale $\xi^*$. To this end, we neglect the loss terms
$\gamma_{\rm C,R,F}$ to obtain, in the case of a Gaussian excitation pulse $P=P_0\ee^{-\left(t/t_0\right)^2}$, an analytical solution
for $\nR$ and $\nF$ in the illuminated area
\begin{align}
\nF &=\frac{\sqrt{\pi}P_0 t_0}{2}  \left[{\rm erf} (t/t_0)+1\right], \\
\nR &\approx \pi t_0^2 \kappa P_0^2 t + C = at + C  \qquad {\rm for}\, t>t_0,
\end{align}
where $a\sim P_0^2$. Initially, the polariton density grows roughly like
\beq
n\approx n_0 \ee^{\frac{1}{2} a\R \left(t-t_{\rm th}\right)^2},
\eeq
where $n_0$ is the initial polariton occupation determined by the Wigner noise and $t_{\rm th}$ is the time when
the critical point is crossed.
Density $n$ saturates at $\nsat$, at the time approximately equal to $\tilde{t}$
\beq
\tilde{t}-t_{\rm th} = \sqrt{\frac{2 \ln \left(\nsat / n_0\right)}{a\R }}.
\eeq
Note that $\tilde{t}-t_{\rm th} \sim P_0^{-1}$ up to a logarithmic factor. We can now estimate 
\beq
\epsilon_{\rm max} \approx \epsilon(\tilde{t}) = \sqrt{2 a \R \ln \left(\nsat - n_0\right) } \sim P_0
\eeq
Since $\xi^* = (\hbar / 2 m^* \epsilon_{\rm max})^{1/2}$ as discussed in the text, we obtain the approximate scaling law
\beq
\label{scaling_law}
N_d \sim L / \xi^* \sim P_0^{1/2}\,.
\eeq

When looking at evolutions at different pumping intensities, we encounter universal behavior
due to the scaling laws for the Bogoliubov frequencies. Since we may write 
$\omega_+(k,t)=\omega_+(k/\xi^*,\alpha(t))=\gamma_0 f(k/\xi^*,(t-t_{\rm th})/\tau)$, where $\tau=t_{\rm max} - t_{\rm th}$
gives the time scale of the growth,
and $\gamma_0\sim\tau^{-1}$ up to a logarithmic factor, we have
\begin{align}
u,v^* &\sim \ee^{-\int_{t_{\rm th}}^{t} i\omega_+(k,t') dt'} = \ee^{-\int_{u_1}^{u_2} i \gamma_0 \tau f(k/\xi^*,u) du} = \nonumber\\
&= g(k/\xi^*,(t-t_{\rm th})/\tau)\,.
\end{align}


\begin{thebibliography}{99}
\bibitem{counting_fluctuations} To determine the number of domains $N_{\rm d}$ we count
the number of zero crossings of the function $f(x)=n(x) - 0.5 n_{\rm max}$, where $n_{\rm max}$
is the maximum of local polariton density over the entire experiment, at the time instant when the number 
of these crossings is the largest. We checked that this method gives reliable estimation of $N_{\rm d}$.
In the disorder-free case, we count domains in single-pulse experiments using the 
exciton density profile $\nR(x)$ instead of the polariton density $n(x)$.
\end{thebibliography}
\end{document}